\documentclass[twocolumn,showpacs,preprintnumbers,amsmath,amssymb]{revtex4}

\usepackage{graphicx}
\usepackage{dcolumn}
\usepackage{bm}
\usepackage{bbold}

\begin{document}

\preprint{APS/123-QED}

\title{Spin projection and spin current density within relativistic electronic transport calculations}

\author{S. Lowitzer}
\author{D. K\"odderitzsch}
\author{H. Ebert}
\affiliation{%
Department Chemie, Physikalische Chemie, Universit\"at M\"unchen, Butenandstr. 5-13, 81377 M\"unchen, Germany\\
}%

\date{\today}

\begin{abstract}
A spin projection scheme is presented which allows the decomposition
of the electric conductivity into two different spin channels within
fully relativistic $ab$ $initio$ transport calculations that account
for the impact of spin-orbit coupling. This is demonstrated by
calculations of the spin-resolved conductivity of Fe$_{1-x}$Cr$_x$ and
Co$_{1-x}$Pt$_x$ disordered alloys on the basis of the corresponding
Kubo-Greenwood equation implemented using the Korringa-Kohn-Rostoker
coherent potential approximation (KKR-CPA) band structure method. In
addition, results for the residual resistivity of diluted Ni-based
alloys are presented that are compared to theoretical and experimental
ones that rely on Mott's two-current model for spin-polarized
systems. The application of the scheme to deal with the spin-orbit
induced spin Hall effect is discussed in addition.
\end{abstract}

\maketitle


During the last years research activities in spintronics increased
very rapidly. The reason for the growing interest in this field is
based on the close connection with fundamental scientific questions as
well as its impact on technology \cite{WAB01,AS09}. Compared to
standard electronics where only the charge of the electrons is used,
spintronics uses the charge of the electrons in combination with the
spin degree of freedom. One of the most exciting effects within
spintronics is the spin Hall effect (SHE) \cite{SCN04,VT06}. The SHE
appears when an electric current flows through a medium with
spin-orbit coupling present, leading to a spin current perpendicular
to the charge current. This effect is even present in non-magnetic
materials as could be demonstrated experimentally e.g. for Pt
\cite{KOS07}. \\For a theoretical investigation of effects like the
SHE it is obviously crucial to have a reliable description for the
spin-dependent transport that accounts for the impact of spin-orbit
coupling in a proper way.  Most investigations in this field were
based on the Pauli equation including spin-orbit coupling explicitly
as a relativistic correction term and representing the spin current
density essentially by a combination of the Pauli spin matrix
$\sigma_z$ with the conventional current density operator
\cite{TKN+08}. Very few investigations have been done so far on the
basis of the Dirac equation using an expression for the spin-current
density, albeit introduced in an ad-hoc manner \cite{GYN05}. In
contrast to these approximate schemes to deal with spin-dependent
transport the approach suggested by \citet{VGW07} supplies a fully
relativistic and coherent description of electronic spin-polarization
and the associated spin-current density. This approach based on the
four-component polarization operator ${\cal T}$ introduced by Bargmann
and Wigner \cite{BW48} leads, in particular, to a corresponding set of
continuity equations. \\ In the present work we introduce spin
projection operators derived from the polarization operator ${\cal
T}$. This allows a decomposition of the conductivity into
contributions from each spin channel within fully relativistic
transport calculations.  Applications on the spin-dependent transport
of various magnetic transition metal alloy systems demonstrate the
flexibility and reliability of the new approach.

Within non-relativistic quantum mechanics the electronic spin can be
described via the well known Pauli matrices $\sigma_i$, specifying the
non-relativistic spin operator ${\bf s}=\frac{\hbar}{2}{\bm
\sigma}$. Due to the fact that the Schr\"odinger Hamiltonian $H_S$
commutes with ${\bf s}$ the projection of the spin, e.g. to the
$z$-axis, is a constant of motion. This is no longer the case within a
scheme that accounts for spin-orbit coupling. The most reliable
approach in this context makes use of electronic structure
calculations on the basis of the Dirac equation.  It turns out that
even in the simplest case of a free electron the Dirac Hamiltonian
does not commute with e.g. $s_z$. However, it is possible to define a
generalized spin operator which commutes with the free electron Dirac
equation and shows all characteristic properties of a spin operator
\cite{FG61,Ros61}.\\ Within the fully relativistic description it is
not possible to decompose the conductivity in a strict sense into
spin-up and spin-down contributions in a simple way. Therefore, one
may use approximative schemes or one can switch to
scalar-relativistic calculations \cite{EVB96,BEV97} to decompose the
conductivity into two different spin channels. The disadvantage of
these two approaches is that approximative schemes work only under
certain circumstances and scalar-relativistic calculations neglect all
scattering events that lead to a spin flip due to the fact that such
calculations neglect spin-orbit coupling.  To avoid such shortcomings
a proper relativistic spin projection operator is necessary. \\
The starting point of our derivation of  suitable relativistic spin
projection operators is based on the four-vector polarization operator
${\cal T}$ which was derived by \citet{BW48}:
\begin{eqnarray}
{\bf T} &=& \beta {\bm \Sigma} - \frac{\gamma_5{\bm \Pi}}{mc}  \\
T_4 &=& i  \frac{{\bm \Sigma}{\bf \cdot {\bm \Pi}}}{mc} \; ,
\end{eqnarray}
with the kinetic momentum ${\bm \Pi}= (\hat{\bf p}+\frac{|e|}{c}{\bf
A})\mathbb{1}_4$ and the canonical momentum $\hat{{\bf p}}$. The
matrices ${\boldsymbol \Sigma}$ are the relativistic Pauli-matrices,
$\beta$ is one of the standard Dirac matrices and \cite{Ros61}:
\begin{equation}
\gamma_5 = \left(  \begin{array}{cc}0&-\mathbb{1}_2 \\-\mathbb{1}_2&0 \end{array}\right) \; .
\end{equation}
The operator ${\cal T}$ can be considered as a generalized spin
operator which commutes with the field free Dirac Hamiltonian
\cite{Ros61}:
\begin{equation}
H^{\rm free} = c {\bm \alpha} {\bf \hat{\cdot p}} + \beta mc^2 \; .
\end{equation}
In addition, the components ${\cal T}_\mu$ are the generators of the
little group that is a subgroup of the group of Lorentz
transformations \cite{FG61}. In comparison to other suggested forms of
polarization operators the operator ${\cal T}$ is gauge invariant
\cite{Ros61} and therefore the appropriate basis for calculations
which include electromagnetic fields.\\ A widely used relativistic
scheme to deal with magnetic solids within spin density functional
theory was introduced by \citet{MV79}. The corresponding Dirac
Hamiltonian:
\begin{equation}
\label{eq:Dirac_B}
H = c {\bm \alpha} {\bf \hat{\cdot p}} + \beta mc^2  + V_{\rm eff} -  \beta {\bm \Sigma}{\bf \cdot B}_{\rm eff} \; ,
\end{equation}
includes an effective scalar potential $V_{\rm eff}$ and an effective
magnetic field ${\bf B}_{\rm eff}$ coupling only to the spin degree of
freedom. For the subsequent discussion we choose ${\bf B}_{\rm eff}=
B(r)\,\hat{\bf e}_z$ as frequently done within electronic structure
calculations. The commutator of ${\cal T}$ and $H$ is non
zero which shows that ${\cal T}$ is no longer a constant of motion.

From ${\cal T}$  corresponding spin projection operators ${\cal
P}^{\pm}$ can be derived by demanding:
\begin{eqnarray}
{\cal P}^+ + {\cal P}^- &=&1 \\
 {\cal P}^+ - {\cal P}^- &=& {\cal T} \; , 
\end{eqnarray}
or equivalently 
\begin{equation}
{\cal P}^{\pm} = \frac{1}{2}(1 \pm {\cal T}) \; . 
\end{equation}
The projection of ${\cal T}$ to a unit vector along the $z$-axis
${\bf n}^T = (0,0,1,0)$ leads to the following expression: 
\begin{equation}
\label{eq:Tn}
{\cal T}\cdot {\bf n}  = \beta \Sigma_z -\frac{\gamma_5 \Pi_z}{mc} .
\end{equation}

Making use of the relation ${\bf B} = \nabla \times {\bf A}$ between
the vector potential ${\bf A}$ and the magnetic field ${\bf B}$, ${\bf
A}$ has only non-zero components in the $xy$-plane if ${\bf B}
\parallel \hat{\bf e}_z$ (see Eq. (\ref{eq:Dirac_B})), i.e. $A_z =
0$. For the spin projection operators this leads to:
\begin{equation}
{\cal P}^{\pm}_z = \frac{1}{2}\left[1 \pm \left(\beta \Sigma_z  -\frac{\gamma_5 \hat{p}_z}{mc}\right)\right] \; . 
\end{equation}

Starting from the polarization operator ${\cal T}$ \citet{VGW07} could
demonstrate that a corresponding spin current density operator is
given by a combination of ${\cal T}$ with the conventional relativistic
electron current density operator: 
\begin{equation}
\label{eq:operator}
\hat{j}_{\mu} = -|e|c\,{\bm \alpha}_{\mu}
\end{equation}
where ${\bm \alpha}_{\mu}$ is one of the standard Dirac matrices
\cite{Ros61}. Accordingly, we get an operator for the spin-projected
current density by combining ${\cal P}^{\pm}_z$ and $\hat{j}_{\mu}$
which leads to ${\cal J}_\mu^{z\pm}={\cal P}^{\pm}_z\hat{j}_{\mu}$.\\
Using ${\cal J}_\mu^{z\pm}$ to represent the observable within Kubo's
linear response formalism one can derive expressions for a
corresponding spin-projected conductivity tensor (the details will be
published elsewhere). Restricting to the symmetric part of the tensor
one arrives at:
\begin{equation}
\label{eq:KG_spin}
\sigma_{\mu\nu}^{z\pm} =  \frac{\hbar}{\pi N\Omega} \, {\rm{Tr}}\,\bigg\langle  {\cal J}_\mu^{z\pm} \, \Im G^+(E_F) \, \hat{j}_{\nu} \, \Im G^+(E_F) \bigg\rangle \;.
\end{equation}
Here $N$ is the number of atomic sites, $\Omega$ the volume per atom,
$\hat{j}_{\mu}$ is the current density operator ($\mu=x,y,z $) and
$\Im G^+(E_F)$ is the imaginary part of the retarded one particle
Green function at the Fermi energy $E_F$.\\ Eq. (\ref{eq:KG_spin}) is
obviously the counter-part to the conventional Kubo-Greenwood equation
\cite{But85} for the spin-integrated conductivity that is
recovered by replacing ${\cal J}_\mu^{z\pm}$ by $\hat{j}_{\mu}$. 

For the determination of $\Im G^+(E_F)$ we use multiple scattering
theory (MST) which is the basis of the KKR band structure method.
Within MST the real space representation of $\Im G^+$ has the
following form \cite{FS80}:
\begin{equation}
\Im G^+({\bf r},{\bf r}',E)= 
 \Im \sum_{\Lambda_1 \Lambda_2}
Z^n_{\Lambda_1}({\bf r}_n,E)
 \tau_{\Lambda_1 \Lambda_2}^{nm}(E)
Z_{\Lambda_2}^{m\times}({\bf r}_m,E) \; ,
\end{equation}
with ${\bf r}={\bf R}_n+{\bf r}_n$ and ${\bf r}'={\bf R}_{m}+{\bf
r}_{m}$.  Using a fully relativistic implementation, the wave
functions $Z^n_{\Lambda}\;(Z^{n\times}_{\Lambda})$ are the regular
right (left) hand side solutions of the Dirac equation within cell
$n$, $\tau^{nm}_{\Lambda \Lambda'}$ is the scattering path operator
and $\Lambda = (\kappa,\mu)$ with $\kappa$ and $\mu$ being the
relativistic spin-orbit and magnetic quantum numbers \cite{Ros61}.
\\The configurational average for a disordered alloy - indicated by
the brackets $\langle ... \rangle$ in Eq. (\ref{eq:KG_spin}) - is taken
by means of the Coherent Potential Approximation (CPA) \cite{But85}.

The scheme outlined above has been implemented by a corresponding
extension of the formalism worked out by \citet{But85} to calculate
the residual resistivity of disordered alloys on the basis of the
Kubo-Greenwood equation. As a first step, the underlying electronic
structure of the investigated disordered alloy systems has been
calculated self-consistently on the basis of local spin density
approximation (LSDA) using the parameterization of \citet{VWN80}. For
the band structure calculations the fully relativistic version of the
Korringa-Kohn-Rostoker (KKR) method \cite{Ebe00} has been used in
combination with the coherent potential approximation (CPA) alloy
theory to account for chemical disorder. The CPA has been exploited in
particular to perform the configurational average when calculating the
conductivity tensor elements and when dealing with the associated
vertex corrections \cite{But85}. These calculations have been done
using a cut-off for the angular momentum expansion at $l_{\rm max}=3$
to ensure convergence for the investigated transition metal systems.

\begin{figure}
 \begin{center}
 \includegraphics[scale=0.73,clip]{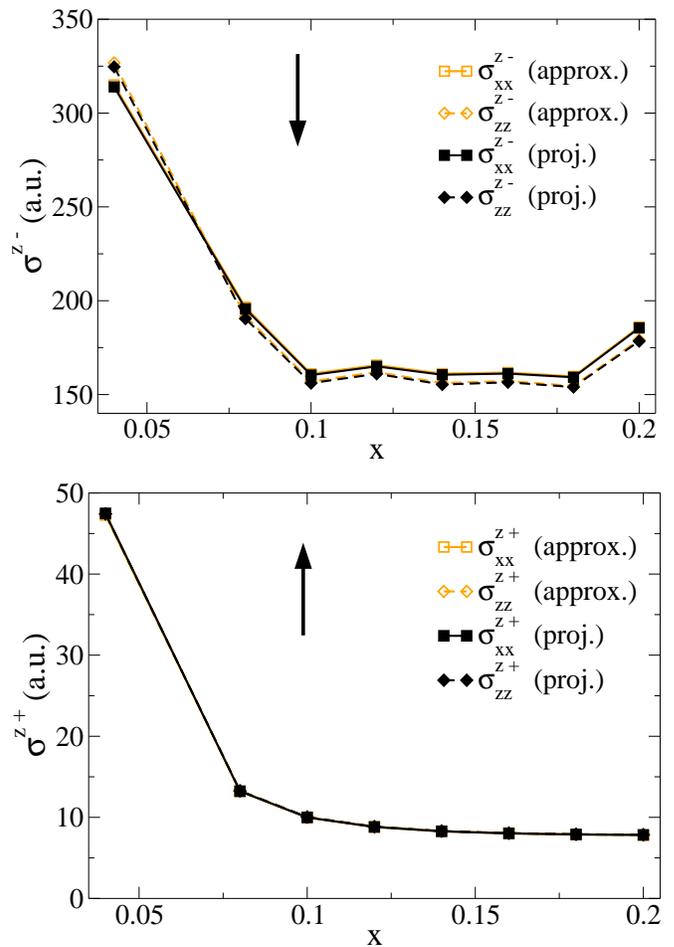}
 \caption{\label{plot:FeCr_op_dow}(Color online) Spin resolved conductivity tensor elements $\sigma_{xx}^{z+(-)}$ and $\sigma_{zz}^{z+(-)}$ of Fe$_{1-x}$Cr$_x$ calculated for the magnetization pointing along the $z$-axis (full symbols). In addition, results are shown that have been obtained using an approximate scheme (open symbols) \cite{PEP+04}.} 
 \end{center}
\end{figure}

As a first application of the presented projection scheme the spin
resolved conductivity of the alloy system Fe$_{1-x}$Cr$_x$ has been
calculated assuming the magnetization to be aligned along the
$z$-axis. The presence of the spin-orbit coupling gives rise to the
anomalous magneto resistance (AMR) with the conductivity tensor
elements $\sigma_{xx}=\sigma_{yy}\neq\sigma_{zz}$ for this
situation. The reduced symmetry is also reflected by the spin
projected conductivities $\sigma_{xx}^{z+(-)}$ and
$\sigma_{zz}^{z+(-)}$, as can be seen in
Fig.~\ref{plot:FeCr_op_dow}.
Obviously, the conductivity is quite different for the two spin
channels. This behavior can be traced back straight forwardly to the
electronic structure of the alloy system around the Fermi energy that
can be represented in a most detailed way in terms of the
spin-projected Bloch spectral function (BSF) \cite{LKE09}. While for
the spin down subsystem there exists a well-defined Fermi surface with
dominant $sp$-character corresponding to a sharp BSF, the spin up
subsystem is primarily of $d$-character that is much more influenced
by the chemical disorder in the system leading to a BSF with rather
washed-out features \cite{LKE09}. As the width of the BSF can be seen
as a measure for the inverse of the electronic lifetime the very
different width found for the two spin subsystems explain the very
different spin-projected conductivities.\\ Fig.~\ref{plot:FeCr_op_dow}
shows in addition results that have been obtained on the basis of an
approximate spin-projection scheme that was suggested recently
\cite{PEP+04}. Within this scheme the matrices occurring in the
Kubo-Greenwood equation for the conductivity are transformed from the
standard relativistic representation (using the quantum numbers
$\Lambda=(\kappa,\mu)$ as labels) to a spin-projected one (using the
quantum numbers $L=(l,m_l,m_s)$ as labels). Suppressing the spin-flip
term of the current density matrix elements $J_{LL'}$ one can easily
split the conductivity into spin-up and spin-down contributions and an
additional spin-flip contribution $\sigma^{z+-}$ that
is related to the spin-off-diagonal elements of the scattering path
operator $\tau$. For 3$d$-elements with a relatively low spin-orbit
coupling it was found that the neglect of spin-off-diagonal elements
of $J_{LL'}$ is well justified and that $\sigma^{z+-}$
is quite small. In fact the spin-projected conductivities
$\sigma_{xx}^{z+(-)}$ and
$\sigma_{zz}^{z+(-)}$ obtained by the approximate
scheme compare very well with the results based on the scheme
presented here (see Fig.~\ref{plot:FeCr_op_dow}).

\begin{figure}
 \begin{center}
 \includegraphics[scale=0.34,clip]{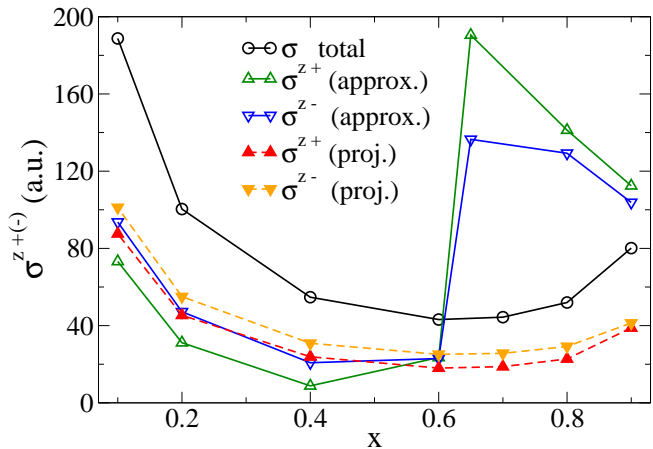}
 \caption{\label{plot:CoPt_up}(Color online) Isotropic spin resolved conductivity $\sigma^{z+(-)}=(2\sigma_{xx}^{z+(-)}+\sigma_{zz}^{z+(-)})/3$ of Co$_{1-x}$Pt$_x$ for the magnetization pointing along the $z$-axis (full symbols). In addition, results are shown that have been obtained using an approximate scheme (open symbols) \cite{PEP+04}.}
 \end{center}
\end{figure}

In order to demonstrate the limitations of the approximate scheme from
Ref.~\cite{PEP+04} the isotropic spin resolved conductivity for
Co$_{1-x}$Pt$_x$ is shown in Fig.~\ref{plot:CoPt_up}.
It turns out, that the approximative scheme fails especially for high
Pt concentrations. This can be attributed to an increased spin-orbit
interaction for which the assumptions on which this scheme is based
are no longer fulfilled.

\begin{figure}
 \begin{center}
 \includegraphics[scale=0.690,clip]{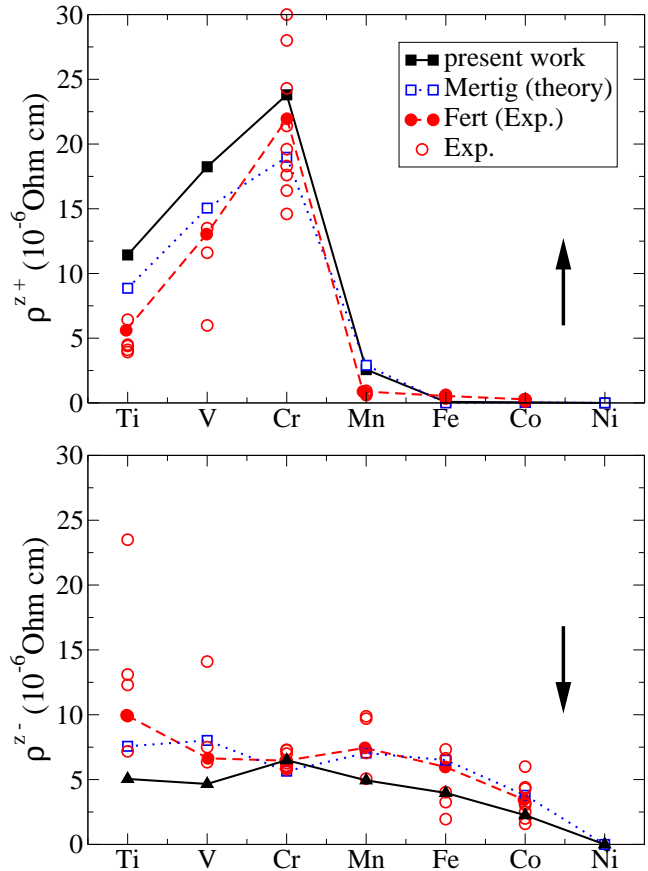}
 \caption{\label{plot:Ni_up}(Color online) Isotropic spin resolved resistivity of Ni with 3$d$ transition metal impurities (1\%) obtained by the present scheme (full squares) compared to theoretical data from \citet{MZD93} (blue squares/dotted line), experimental data from \citet{FER08} (full red circles/dashed line) and other experimental data (see Ref. \onlinecite{DM74}, open red circles). The top and the bottom panel show the data for spin-up and for spin-down, respectively.}
 \end{center}
\end{figure}

As another application of the presented scheme results for diluted
Ni-based alloys with $x_{\rm Ni}=0.99$ are shown in
Fig.~\ref{plot:Ni_up} in terms of the isotropic residual resistivities
$\rho^{z+(-)} =
[(2\,\sigma_{xx}^{z+(-)}+\sigma_{zz}^{z+(-)})/3]^{-1}$.
As one notes, the resistivity for the two spin channels show a rather
different variation with the atomic number of the impurities. This can
be traced back again to the spin-projected electronic structure of Ni
at the Fermi level and the position of the impurity $d$-states
\cite{MZD93}. In Fig.~\ref{plot:Ni_up} the results of calculations by
\citet{MZD93} have been added, that were done in a scalar-relativistic
way - i.e. ignoring spin-orbit coupling - on the basis of the
Boltzmann-formalism and by making use of the two-current model. In
spite of the various differences between this approach and the
presented scheme, the resulting spin-projected resistivities agree
fairly well. This also holds concerning corresponding experimental
data that have been deduced from measurements relying on the
two-current model.

In summary, a scheme for a spin projection within transport
calculations on the basis of the Kubo formalism has been
presented. The applications presented were restricted to the diagonal
elements of the corresponding conductivity tensor described by a
Kubo-Greenwood-like equation. Results obtained for the disordered
alloy systems Fe$_{1-x}$Cr$_x$, Co$_{1-x}$Pt$_x$ and diluted Ni-based
alloys were compared to results based on an alternative but
approximate projection scheme and theoretical as well as experimental
data based on the two current model. The good agreement found for the
investigated systems ensures the consistency and reliability of the
presented scheme. Accordingly, this is expected to hold also when
dealing with spin-projected off-diagonal conductivities as
e.g. $\sigma_{xy}^{z+(-)}$ on the basis of
Kubo-St\v{r}eda-like equations. This will give access in particular to
the spin-projected Hall conductivity in magnetic materials as well as
to the spin Hall conductivity in non-magnetic materials. Work along
this line is in progress.

The authors would like to thank the DFG for financial support within
the SFB 689 ``Spinph\"anomene in reduzierten Dimensionen''. D.K. in
addition acknowledges support from the DFG priority program SPP 1145
``Modern and universal first-principles methods for many-electron
systems in chemistry and physics''.
\newpage 

\end{document}